\documentclass[draftclsnofoot, onecolumn,12pt]{IEEEtran}
\IEEEoverridecommandlockouts
\usepackage{cite}
\usepackage{amsmath,amssymb,amsfonts, mathtools}
\usepackage{graphicx}
\usepackage{subfigure}
\usepackage{textcomp}
\usepackage{xcolor}
\usepackage{soul}
\usepackage{bm}
\usepackage[noend]{algpseudocode}
\usepackage{lipsum}
\usepackage{algorithmicx,algorithm}
\usepackage{multirow}

\usepackage{tikz}
\usetikzlibrary{patterns}

\newtheorem{Rem}{Remark}

\def\BibTeX{{\rm B\kern-.05em{\sc i\kern-.025em b}\kern-.08em
    T\kern-.1667em\lower.7ex\hbox{E}\kern-.125emX}}

\graphicspath{{./figure/}}

\begin{document}

\title{Low-complexity Joint Beamforming for RIS-Aided Multi-User Downlink over Correlated Channels}

\author{Yu-Tse Wu
        and Kuang-Hao Liu,~\IEEEmembership{Member,~IEEE}
\thanks{Y.-T.~Wu is with the Department of Electrical Engineering,
National Cheng Kung Univrsity, Tainan, Taiwan 701.}
\thanks{(E-mail: {\tt brad7787711911@gmail.com})}
\thanks{K.-H.~Liu is with the Institute of Communications Engineering, National Tsing Hua University, Hsinchu, Taiwan 300.}
\thanks{(E-mail: {\tt khliu@ee.nthu.edu.tw})}
} 


\maketitle
\begin{abstract}
This paper considers the reconfigurable intelligent surface (RIS)-assisted multi-user communications, where an RIS is used to assist the base station (BS) for serving multiple users. The RIS consisting of passive reflecting elements can manipulate the reflected direction of the incoming electromagnetic waves by adjusting the phase shifts of the reflecting elements. Alternating optimization (AO) based approach is commonly used to determine the phase shifts of the RIS elements. While AO-based approaches have shown significant gain of RIS, the complexity is quite high due to the coupled structure of the cascade channel from the BS through RIS to the user. In addition, the sub-wavelength structure of the RIS introduces spatial correlation that may cause strong interference to users. To handle severe multi-user interference over correlated channels, we consider adaptive user grouping previously proposed for massive mutli-input and multi-output (MIMO) systems and propose two low-complexity beamforming design methods, depending on whether the grouping result is taken into account. Simulation results demonstrate the superior sum rate achieved by the proposed methods than that without user grouping. Besides, the proposed methods can perform similarly to the AO-based approach but with much lower complexity. 
\end{abstract}

\begin{IEEEkeywords}
Reconfigurable intelligent surface, user grouping, correlated channel, maximum sum rate.
\end{IEEEkeywords}

\section{Introduction}
Recently, reconfigurable intelligent surface (RIS) has been received much attention as an energy-efficient technology for future wireless communications~\cite{Wu2019,Tang2020,Liaskos2018,Basar2019,Wu2020}. An RIS is a meta-surface consisting of a large number of low-cost, energy-efficient, and passive reflecting elements. The phase of each element can be electronically controlled to reflect the radio signals in a desired manner, such as interference suppression ~\cite{Wu2020}, signal power enhancement ~\cite{Wu2018}, and sum rate maximization ~\cite{Yang2020,Chen2019}. 

The merits of RIS stimulate recent interest on RIS-assisted multi-user communications where the passive phase shifts at RIS and the active beamforming at the BS are jointly optimized. In~\cite{Wu2020a}, practical RISs are considered where the passive phase shifts can only take discrete values. The authors formulate the optimal beamforming design problem aiming to minimize the BS transmission power subject to the user signal-to-interference-plus-noise ratio (SINR) and the RIS discrete phase shift constraints. Since the BS beamforming and RIS phase shit coefficients are jointly optimized using alternating optimization (AO), the complexity is very high and does not scale well as the numbers of users and RIS elements increase. To permit practical implementation, a suboptimal solution was proposed where the BS beamfomring is determined based on the zero-forcing (ZF) principle for a fixed RIS phase shift matrix. Then a one-dimensional search is used to find the best beamforming pair. The discrete RIS phase shift design for power minimization problem is also considered in the context of non-orthogonal multiple access (NOMA)~\cite{Zheng2020a}. Different from~\cite{Wu2020a},~\cite{Zheng2020a} focuses on BS power allocation and RIS phase shift design, whose joint optimization is NP hard. The authors decompose the joint optimization problem into two subproblems where the power allocation subproblem is solved for the given RIS phase shifts. To reduce the complexity of element-wise phase shift optimization, the RIS elements are divided into several sub-surfaces each sharing the same phase shift coefficient that is alternately optimized. Another attempt to solve the challenging optimization of joint BS beamforming and RIS phase shifts is reported in~\cite{Ma2021}, where a novel problem decomposition based on fractional programming (FP) was proposed. The machine learning approach is also applied to the joint beamforming design for RIS-assisted multi-user communications~\cite{Huang2020}. 

In this work, we focus on low-complexity methods to solve the joint beamforming design when a single RIS is deployed to serve multiple users. While some low-complexity methods have been proposed~\cite{Wu2020a, Zheng2020a, Ma2021}, they all need certain iterations to obtain converged results and the computational complexity increases with the number of users and RIS elements. In the proposed methods, the complexity only grows with the number of RIS phase-shift discrete levels yet its sum-rate performance is close to that using AO-based refinement. Besides, prior work commonly assumes independent fading channels that may not always hold when the propagation channels have less scattering and the RIS elements are densely deployed with sub-wavelength periodicity~\cite{Bang2018, Chien}. With spatial correlated channels, the mutual interference among users may be strong that further challenges the joint BS beamforming and RIS phase shift design. In view of this difficulty, user grouping is introduced to remedy the design challenge. With user grouping, users are separated into different groups each with a small number of users and low spatial correlation. Consequently, the BS beamforming can be easily designed and then optimized jointly with the RIS phase shifts. We note that the impact of spatial correlated channels is studied in~\cite{Zhao2021,Chien}, demonstrating dramatic sum rate loss in the presence of spatial correlation. We also take into account the impact of the overhead for tuning the RIS and propose two transmission protocols that capture the tradeoff between beamforming accuracy and RIS configuration overhead. Prior work has studied the training overhead for acquiring channel state information~\cite{Zhou2020}.

The remainder of the paper is organized as follows. Sec.~\ref{sec: model} explains the considered RIS system model. In Sec.~\ref{sec: proposed}, we present two low-complexity methods to determine the RIS phase shifts. The correlation-based grouping algorithm is introduced in Sec.~\ref{sec: grouping}. Numerical results and discussions are provided in Sec.~\ref{sec: results}, and finally, Sec.~\ref{sec: conclusion} summarizes this paper.

\section{System Model}\label{sec: model}
\begin{figure}[!t]
\centering
{
\includegraphics[width=1\linewidth]{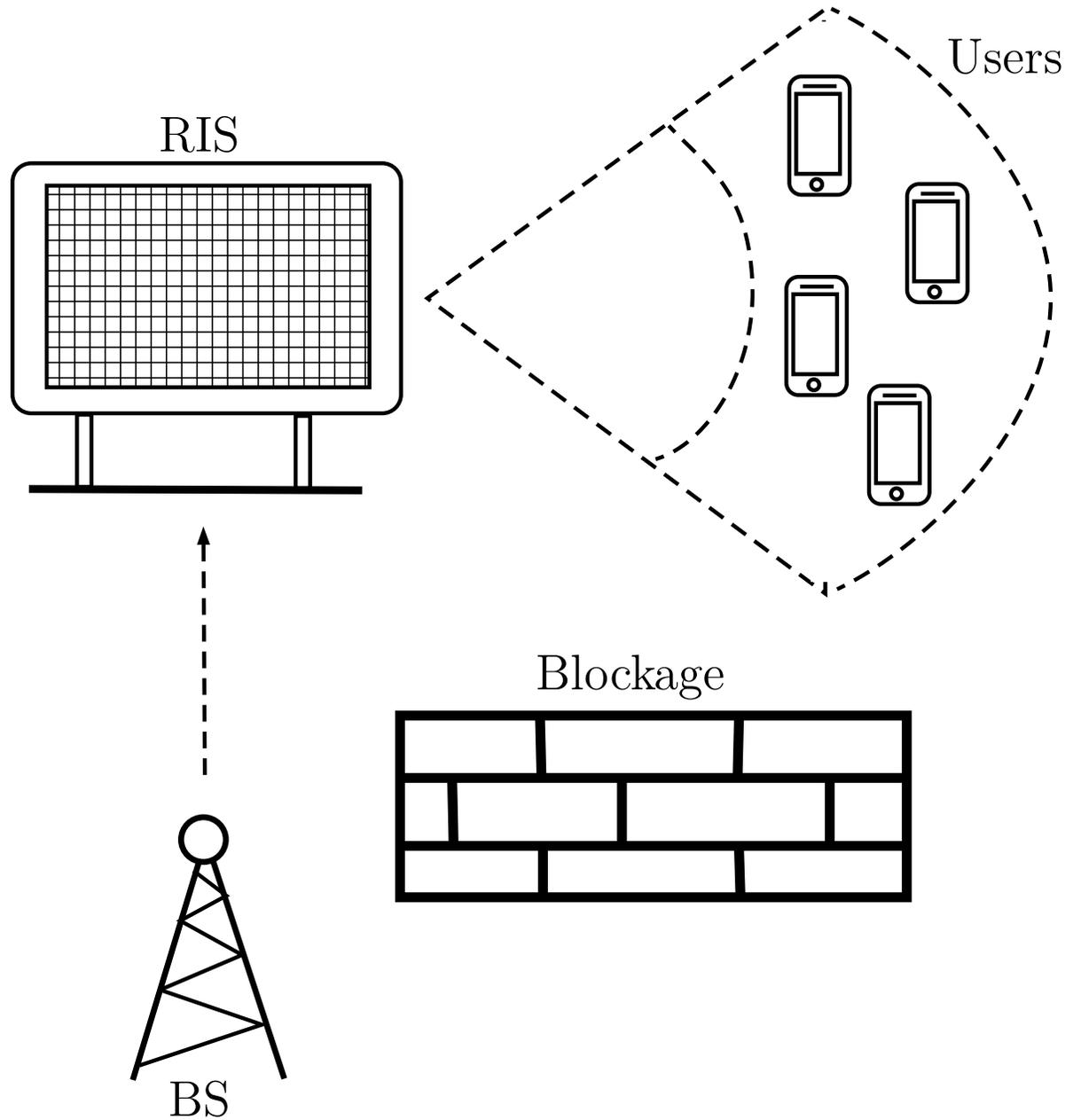}
}
\caption{Illustration of the RIS-assisted multi-user communication scenario.}
\label{fig: system-model}
\end{figure}

We consider an RIS-assisted downlink system with $K$ single-antenna users served by a BS, which is equipped with $M$ antennas, as shown in Fig.~\ref{fig: system-model}. The RIS is composed of $N$ passive reflecting elements. Let $\mathbf{H}_{t} \in \mathbb{C}^{N \times M} $ denote the baseband equivalent channel from the BS to the RIS and $ \mathbf{h}_{r,k}^{H} \in \mathbb{C}^{1 \times N} $ denote the channel from the RIS to the $k$th user, $k = 1,\cdots,K$. The reflection coefficients of the RIS are denoted by $ \bm{\theta} \in \mathbb{C}^{1 \times N} \triangleq [\nu_{1}e^{j\phi_{1}},\nu_{2}e^{j\phi_{2}},\cdots,\nu_{N}e^{j\phi_{N}}] $, where $\phi_{n} $ is the phase shift subject to the discrete reflection constraint with discrete level $L$~\cite{Zheng2020a,Wu2020a} and $ \nu_{n} \in \{0,1\} $ is the reflection amplitude of the $n$th element. In this work, we assume $\nu_n = 1$, $\forall n=1,\cdots,N $ to maximize the signal power reflected by the RIS.

%
%
%
%

A scheduling cycle of a fixed duration is considered and it is divided into equal-length time slots. To mitigate strong multi-user interference, the adaptive user grouping algorithm~\cite{Alkhaled2016} proposed for massive MIMO is applied that separates correlated users in different groups. Each group is assigned with dedicated slots while the users in the same group share the same slot. To maintain a certain fairness, the number of time slots allocated to each group per scheduling cycle is proportional to the number of users in a group. Detailed grouping mechanism will be given in Sec.~\ref{sec: grouping}. The considered user grouping can greatly simplify the beamforming design because the user channels in the same group have a smaller correlation and dimension than the overall user channels. In this case, sub-optimal beamforming design can often yield good performance with minor performance loss than the sophisticated optimal design.


Denote $\mathcal{G}_g$ the $g$th user group and there are total of $N_G$ groups. With the aid of the RIS, the received signal of the $k$th user in the $g$th group is given by

\begin{flalign}\label{eq: recv-signal}
\qquad y_k^{\mathcal{G}_g} =
& \sqrt{P_{t,k}} \mathbf{h}_{r,k}^{H} \boldsymbol{\Theta} \mathbf{H}_{t} \mathbf{w}_{k} x_k && \nonumber \\
&~ +\sum_{j \in \mathcal{G}_g , j \neq k} \sqrt{P_{t,j}} \mathbf{h}_{r,k}^{H} \boldsymbol{\Theta} \mathbf{H}_{t} \mathbf{w}_{j} x_j + n_k
\end{flalign}
where $P_{t,k} = P_{t}/|\mathcal{G}_g|$ is the transmit power from the BS to user $k$ while $P_t$ is the total transmit power from the BS, $\boldsymbol{\Theta} = \text{diag} (\bm{\theta})$ is a diagonal matrix for determining the RIS phase shifts, $\mathbf{w}_k \in \mathbb{C}^{M \times 1}$ is the $k$th column of the beamforming matrix $\mathbf{W}$ at the BS, $ x_k $ is the symbol to be transmitted to the $k$th user with $ \mathbb{E}\{|x_k|^2\} = 1$, and $n_k$ is the additive white Gaussian noise (AWGN) at the $k$th user with power equals to $\sigma^2_n$ dBm/Hz. In (\ref{eq: recv-signal}), the first term is the desired signal of the $k$th user and the second term is the multi-user interference from other users' signals in the same group. Accordingly, the instantaneous signal-to-noise-plus-interference ratio (SINR) of the $k$th user in the $g$th group can be presented as
\begin{equation}
\text{SINR}_k^{\mathcal{G}_g} = \frac{P_{t,k} | \mathbf{h}_{r,k}^{H} \boldsymbol{\Theta} \mathbf{H}_{t} \mathbf{w}_{k} |^2}
{ \sum_{j \in \mathcal{G}_g , j \neq k} P_{t,j} |\mathbf{h}_{r,k}^{H} \boldsymbol{\Theta} \mathbf{H}_{t} \mathbf{w}_{j}|^2
+ B\sigma_n^2}
\end{equation}
where $B$ denotes the system bandwidth. Given $\text{SINR}_k^{\mathcal{G}_g}$, the achievable rate of the $k$th user in the $g$th group can be evaluated as
\begin{equation}
R_{k}^{\mathcal{G}_g} = B \log_2 (1+\text{SINR}_k^{\mathcal{G}_g}).
\end{equation}


\section{Low-complexity method for designing reflection coefficients}\label{sec: proposed}

Generally, $\boldsymbol{\Theta}$ should be designed jointly with $\mathbf{W}$ to maximize a certain target function subject to the uni-modulus constraint of the reflection coefficients, i.e., $| e^{j \phi_n} |^2 = 1$. Thus the joint beamforming design problem for the RIS-assisted communications can be formulated as
\begin{flalign}
(\mathtt{P}) \qquad \qquad \qquad \qquad
& \max_{\mathbf{W}, \boldsymbol{\Theta}} J(\mathbf{W}, \boldsymbol{\Theta}) && \nonumber \\
\text{s.t.}
&~ |e^{j\phi_{n}}|^2 = 1, \forall~{n} = 1,\cdots,N
\end{flalign}
where $J(\mathbf{W}, \boldsymbol{\Theta})$ represents the target function. For the downlink transmissions, one commonly considered target function is the sum rate~\cite{Ma2021,Zhao2021}, given by $\sum_{k=1}^K R_k^{\mathcal{G}_g}$. In this case, the target function is non-convex and the optimization variables $\mathbf{W}$ and $\boldsymbol{\Theta}$ are coupled. Instead of solving $(\mathtt{P})$ using AO-based approach, we propose two low-complexity methods to determine $\mathbf{W}$ and $\boldsymbol{\Theta}$. 

esults in the maximal received power and is jointly convex in $\mathbf{W}$ and $\boldsymbol{\Theta}$. However, $\mathbf{W}$ and $\boldsymbol{\Theta}$ are coupled in $J(\mathbf{W}, \boldsymbol{\Theta})$ and thus solving $(\mathtt{P})$ remains difficult. Most of the related work solves $(\mathtt{P})$ using the alternating optimization (AO) approach.  


The reflecting elements on the RIS can be considered as the passive antenna elements. By adjusting the phase shift of each reflecting element, RIS can concentrate the reflected signal to a desired direction. Accordingly, we can construct multiple candidate reflection coefficient matrices such that each of them creates a spatial beam toward a certain direction, which is known as beam steering in array signal processing. For example, the steering vector of the uniform linear array with $N$ elements is given by $[1 \; e^{-j \psi} \; \cdots \; e^{-j (N-1) \psi}]^T$ where $\psi=\pi \sin(\theta)$ is the constant phase difference between two adjacent elements when their distance is half of the wavelength. Suppose that the $l$th candidate reflection coefficient matrix $\boldsymbol{\Theta}_l = \text{diag}(\bm{\theta}_l)$ impinges the reflected signal to angle $\vartheta_l=2\pi l/L$ for $l=0, \cdots, L-1$. Then $\bm{\theta}_l$ can be found as
\begin{equation}\label{eq: ris-phase-vector}
\bm{\theta}_{l} = \left[ e^{-j\pi \cdot 0 \cdot \sin (\vartheta_l)},\cdots, e^{-j\pi(N-1) \sin (\vartheta_l)} \right].
\end{equation}
Clearly, $\bm{\theta}_{l}$ in (\ref{eq: ris-phase-vector}) satisfies the union-modulus constraint. Besides, the union of these spatial beams covers the angular interval of $2\pi$. 

For each $\boldsymbol{\Theta}_l$, one can find the corresponding BS beamforming matrix $\mathbf{W}_l$ to meet a desired target, for example, signal enhancement or interference cancellation. Suppose a user grouping mechanism is in place such that users with correlated channels care separated into different groups, the interference among users within the same group is expected to be small. Thus, designing $\mathbf{W}_l$ for signal enhancement is more beneficial than interference cancellation. In this work, $\mathbf{W}_l$ is determined following the maximum ratio transmission (MRT) principle as given by
\begin{equation}\label{eq: bs-beamforming}
\mathbf{W}_l^{\mathcal{G}_g} = \frac{ (\mathbf{H}_r^{\mathcal{G}_g} \boldsymbol{\Theta}_l \mathbf{H}_{t})^H}{\sqrt{\text{Trace} ((\mathbf{H}_r^{\mathcal{G}_g} \boldsymbol{\Theta}_
l \mathbf{H}_{t})(\mathbf{H}_r^{\mathcal{G}_g} \boldsymbol{\Theta}_l \mathbf{H}_{t})^H)}}
\end{equation}
where $\mathbf{H}_r^{\mathcal{G}_g} \in \mathbb{C}^{|\mathcal{G}_g| \times N}$ is the reflecting channel matrix which is composed of the channels from the RIS to the users in the $g$th group. 
For each group, the beamforming pair that offers the highest target function value is considered as the solution for $(\mathtt{P})$, which can be expressed as
\begin{equation}\label{eq: beamforming-pair-dependent}
(\mathbf{W}^*,\boldsymbol{\Theta}^*)^{\mathcal{G}_g} = \arg~\max_{\mathbf{W}_l^{\mathcal{G}_g}, \boldsymbol{\Theta}_l^{\mathcal{G}_g}} J(\mathbf{W}_l^{\mathcal{G}_g}, \boldsymbol{\Theta}_l^{\mathcal{G}_g}).
\end{equation}
With $N_G$ groups, the RIS needs to configure the phase shifts $N_G$ times. In practice, phase shift adjustment is often achieved by tuning the structure of the unit cell plane on the RIS with the speed on the order of milliseconds~\cite{Bang2018}. Compared with the slot length in emerging wireless systems, e.g., 5G NR, ranging from 0.0625 ms$\sim$1ms~\cite{TS38211}, the time overhead for RIS phase shift configuration should be considered when evaluating the achieved performance of the beamforming scheme. To this end, denote $t_p$ the proportion of time per scheduling cycle consumed for configuring RIS phase shifts. Since the beamforming pair in (\ref{eq: beamforming-pair-dependent}) is designed for individual group, it is referred to as the \textit{group-based} design with the target function given by 
\begin{equation}\label{eq: rate-group-dependent}
J^{\text{G}} = \sum_{g=1}^{N_G} \left(\frac{|\mathcal{G}_g|}{K} - t_p \right) \sum_{k \in \mathcal{G}_g} R_{k}^{\mathcal{G}_g}
\end{equation}
where $|\mathcal{G}_g|$ represents the number of users in the $g$th group. Thus the factor $|\mathcal{G}_g|/K$ accounts for the proportion of time allocated to the users in the $g$th group. It should be noted that $|\mathcal{G}_g|/K$ must be no less than $t_p$ to ensure enough time for RIS phase shift configuration.

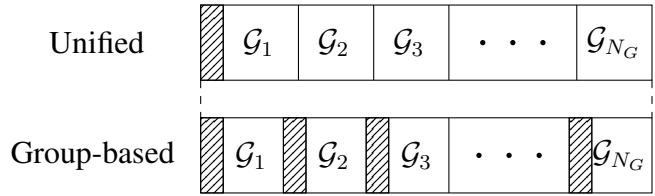
\begin{figure}[!t]
\begin{tikzpicture}
\coordinate [label=left:\textcolor{black}{Unified}] (A) at (-0.6,2);
\coordinate [label=left:\textcolor{black}{Group-based}] (B) at (-0.2,0.5);
\coordinate [label=left:\textcolor{black}{$\mathcal{G}_1$}] (C) at (1.125,2);
\coordinate [label=left:\textcolor{black}{$\mathcal{G}_2$}] (D) at (2.125,2);
\coordinate [label=left:\textcolor{black}{$\mathcal{G}_3$}] (E) at (3.125,2);
\coordinate [label=left:\textcolor{black}{$\mathcal{G}_{N_G}$}] (F) at (6,2);
\draw (0,1.5) -- (6,1.5) -- (6,2.5) -- (0,2.5) -- (0,1.5);
\draw (0.3,1.5) -- (0.3,2.5);
\draw (1.3,1.5) -- (1.3,2.5);
\draw (2.3,1.5) -- (2.3,2.5);
\draw (3.3,1.5) -- (3.3,2.5);
\draw (5,1.5) -- (5,2.5);
\fill (3.75,2) circle (1pt);
\fill (4.15,2) circle (1pt);
\fill (4.55,2) circle (1pt);
\draw [pattern=north east lines, pattern color=black] (0,1.5) rectangle (0.3,2.5);

\coordinate [label=left:\textcolor{black}{$\mathcal{G}_1$}] (F) at (1.025,0.5);
\coordinate [label=left:\textcolor{black}{$\mathcal{G}_2$}] (G) at (2.125,0.5);
\coordinate [label=left:\textcolor{black}{$\mathcal{G}_3$}] (H) at (3.225,0.5);
\coordinate [label=left:\textcolor{black}{$\mathcal{G}_{N_G}$}] (I) at (6.075,0.5);
\draw (0,0) -- (6,0) -- (6,1) -- (0,1) -- (0,0);
\draw (0.3,0) -- (0.3,1);
\draw (1.1,0) -- (1.1,1);
\draw (1.4,0) -- (1.4,1);
\draw (2.2,0) -- (2.2,1);
\draw (2.5,0) -- (2.5,1);
\draw (3.3,0) -- (3.3,1);
\draw (4.9,0) -- (4.9,1);
\draw (5.2,0) -- (5.2,1);
\fill (3.7,0.5) circle (1pt);
\fill (4.1,0.5) circle (1pt);
\fill (4.5,0.5) circle (1pt);
\draw [pattern=north east lines, pattern color=black] (0,0) rectangle (0.3,1);
\draw [pattern=north east lines, pattern color=black] (1.1,0) rectangle (1.4,1);
\draw [pattern=north east lines, pattern color=black] (2.2,0) rectangle (2.5,1);
\draw [pattern=north east lines, pattern color=black] (4.9,0) rectangle (5.2,1);
\draw [dashed] (0,1.5) -- (0,1);
\draw [dashed] (6,1.5) -- (6,1);
\end{tikzpicture}
\caption{Illustration of the scheduling cycle for the unified and the group-based beamforming design.}
\label{fig: unified-vs-based}
\end{figure}

As indicated in (\ref{eq: rate-group-dependent}), the time overhead for phase shift configuration increases with the number of groups $N_G	$. The overhead can be reduced by performing phase configuration only once per scheduling cycle. In this case, the BS beamforming matrix is determined based on the overall user channels as given by
\begin{equation}\label{eq: bs-beamforming-independent}
\mathbf{W}_l = \frac{ (\mathbf{H}_r \boldsymbol{\Theta}_l \mathbf{H}_{t})^H}{\sqrt{\text{Trace} ((\mathbf{H}_r \boldsymbol{\Theta}_
l \mathbf{H}_{t})(\mathbf{H}_r \boldsymbol{\Theta}_l \mathbf{H}_{t})^H)}}
\end{equation}
where  $\mathbf{H}_r \in \mathbb{C}^{K \times N}$ is the reflecting channel matrix whose $k$th row is $\mathbf{h}_{r,k}^{H}$. The beamfoming pair is thus found by solving the following problem.
\begin{equation}\label{eq: beamforming-pair-independent}
(\mathbf{W}^*, \boldsymbol{\Theta}^*) = \arg~\max_{\mathbf{W}_l, \boldsymbol{\Theta}_l} J(\mathbf{W}_l, \boldsymbol{\Theta}_l).
\end{equation}
Since the beamforming pair obtained from (\ref{eq: beamforming-pair-independent}) is fixed over the entire scheduling cycle, (\ref{eq: beamforming-pair-independent}) is referred to as the \textit{unified design}. For the users in the $g$th group, the BS beamforming matrix is the collection of the $k$th column of $\mathbf{W}^*$, $\forall~k \in \mathcal{G}_g$. The target function for the unified design is given by
\begin{equation}
J^{\text{U}} = (1-t_p)\sum_{g=1}^{N_G} \frac{|\mathcal{G}_g|}{K} \sum_{k \in \mathcal{G}_g} R_{k}^{\mathcal{G}_g}.
\end{equation}
Fig.~\ref{fig: unified-vs-based} shows the scheduling cycle for the unified and the group-based beamforming design, where the shaded zone indicates the period for phase shift configuration. Clearly, the unified design saves the overhead for phase shift configuration. Its performance will be compared with the group-based design in Sec.~\ref{sec: results}. We note that $\mathbf{W}_l$ in (\ref{eq: bs-beamforming-independent}) may be determined using the ZF principle to null out the multi-user interference at the BS side. However, ZF performs poor when $\mathbf{H}_r \boldsymbol{\Theta}_l \mathbf{H}_t$ is rank-deficient, the problem caused by spatially correlated channels.



%




\begin{table}
\centering
\caption{Complexity comparison}
\label{tab: complexity}
\begin{tabular}{ c|l }
\hline
\textbf{Algorithm} & \textbf{Complexity} \\
\hline
Proposed algorithm & Unified: $\mathcal{O}(L)$ \\
\hline
                   & Group-based: $\mathcal{O}(L N_G)$ \\
\hline
AO-based refinement  & \cite{Wu2020a}: $\mathcal{O}(I_1L(K^3+K^2 M + KMN))$ \\
\hline
                     & \cite{Zheng2020a}: $\mathcal{O}((B+I_2)SL)$ \\
\hline
FP-based algorithm   & \cite{Ma2021}: $\mathcal{O}((N+MK+3K)I_3)$ \\
\hline
\end{tabular}
\end{table}

Table~\ref{tab: complexity} lists the complexity of the proposed methods and some existing algorithms aiming to low-complexity design for RIS phase shifts, where $I_i$ with $i=1, 2, 3$ is the number of iterations for different algorithms. The notation $B$ represents the number of quantization levels between 0 and 1 used in~\cite{Zheng2020a}. In addition, $S$ denotes the number of sub-surfaces. It is revealed the complexity of the unified design only increases with the discrete levels of RIS phase shifts. The group-based design incurs a higher complexity depending on the number of user groups but its complexity is much less than existing algorithms.

\section{Correlation-based Grouping Algorithm}\label{sec: grouping}

The spatial correlation between two or more users in RIS-assisted system introduces severe multi-user interference that in turn degrades the sum rate.  Since the end-to-end channel characteristics in the RIS system are altered by the reflection coefficient matrix $\boldsymbol{\Theta}$, the correlation coefficient between two different users should take into account the cascade channels as given by
\begin{equation}\label{eq: corr-coeff}
\rho_{i,j} = \frac{|(\mathbf{h}_{r,i}^{H} \boldsymbol{\Theta} \mathbf{H}_{t})\cdot(\mathbf{h}_{r,j}^{H} \boldsymbol{\Theta} \mathbf{H}_{t})^H|}{\Vert \mathbf{h}_{r,i}^{H} \boldsymbol{\Theta} \mathbf{H}_{t} \Vert \cdot \Vert \mathbf{h}_{r,j}^{H} \boldsymbol{\Theta} \mathbf{H}_{t} \Vert},~0 \leq \rho_{i,j} \leq 1.
\end{equation}
where $\mathbf{h}_{r,i}^{H} \boldsymbol{\Theta} \mathbf{H}_{t}$ and $\mathbf{h}_{r,j}^{H} \boldsymbol{\Theta} \mathbf{H}_{t}$ are the channel vectors of the $i$th and $j$th user, respectively, through a common RIS. A higher $\rho_{i,j}$ implies the channel vectors of the $i$th and $j$th user are more correlated. 

\addtolength{\topmargin}{0.01in}


Following the idea in~\cite{Alkhaled2016}, we employ user grouping based on the correlation coefficient in (\ref{eq: corr-coeff}) to mitigate strong multi-user interference in the RIS system. For the readers' convenience, we briefly explain the adaptive user grouping algorithm proposed in~\cite{Alkhaled2016}. First, any two users with their channel correlation coefficient $\rho_{i,j}$ higher than a threshold $\eta$, which we call the \emph{grouping threshold}, is separated into two groups. This procedure is repeated until all users are examined. For those users with $\rho_{i,j} \leq \eta$, they are assigned to the group with the minimum summed correlation coefficients to avoid high interference within the same group. Algorithm~\ref{alg: user grouping} summarizes how the beamforming pair and the user groups are determined.

\begin{algorithm}[t]
\caption{Beamforming Design and Grouping Algorithm}
\label{alg: user grouping}
\hspace*{\algorithmicindent} \textbf{Input}: $U=\{1,2, \cdots, K\},\mathbf{H}_t,\mathbf{H}_r,\eta$
\begin{algorithmic}
\State \textbf{Stage 1 : Decide the $\boldsymbol{\Theta}$ and $\mathbf{W}$}
\For   {each $l \in [0,1,\cdots,L-1]$}
\State Compute $\boldsymbol{\Theta}_l$ by (\ref{eq: ris-phase-vector}) and $\mathbf{W}_l$ by (\ref{eq: bs-beamforming}) or (\ref{eq: bs-beamforming-independent}).
\EndFor
\State Select the optimal beamforming pair $\mathbf{W}^*$ and $\boldsymbol{\Theta}^*$ by
\State solving (\ref{eq: beamforming-pair-dependent}) or (\ref{eq: beamforming-pair-independent}).
\\
\State \textbf{Stage 2 : Group the users}
\State Find $\rho_{i,j}$ using (\ref{eq: corr-coeff}), $\forall~i \neq j \in U$.
\State 1) Separate the high correlated users
\Repeat
\If{$\rho_{i,j} > \eta, \forall~i \neq j \in U$}
\State Split users $i$ and $j$ into two different groups.
\State $U=U\backslash\{i,j\}$
\EndIf
\Until $\rho_{i,j} \leq \eta, \forall~i \neq j \in U$.
\State 2) Append the rest users
\For   {$k \in U$}
\State Calculate $\sum_{j \in \mathcal{G}_g}{\rho_{k,j}}$ for all existing groups $\mathcal{G}_g$ 
\State and append $k$ to the minimum one.
\EndFor
\end{algorithmic}
\end{algorithm}

\begin{Rem}\label{rem: correlation}
The correlation threshold $\eta$ determines the intensity of user grouping. When $\eta$ is small, it is more likely to separate users into different groups and vice versa. For a fixed $\eta$, whether two arbitrary users are separated into different groups depends on the number of RIS elements. With a large $N$, $\rho_{i,j}$ tends to be smaller and thus the condition $\rho_{i,j} > \eta$ is less likely to be satisfied. This results in a few groups each with more users. As $N \rightarrow \infty$, most users will be in the same group. Since the users within the same group will share the identical resource, the sum rate performance is dominated by the multi-user interference. On the contrary, the sum rate performance for a small $N$ is limited by the resource penalty due to many groups created to separate correlated users.
\end{Rem}

\section{Simulation Results}\label{sec: results}
Simulation results are presented to evaluate the performance of the proposed beamforming design along with user grouping. In simulations, the channel between the BS and RIS is modeled by the Rician channel model as given by~\cite{Wu2019}
\begin{equation*}
\mathbf{H}_t = \sqrt {C_0 \left(\tfrac{d_{BR}}{d_0}\right)^{-\alpha_{BR}}} \Bigl(\sqrt{\tfrac{\beta}{\beta+1}} \mathbf{H}_t^{\text{LoS}} + \sqrt{\tfrac{1}{\beta+1}} \mathbf{H}_t^{\text{NLoS}}\Bigr)
\end{equation*}
where $C_0$ is the path loss at the reference distance $d_0 = 1$m, $d_{BR}=50$m denotes the distance from the BS to the RIS, $\alpha_{BR} = 2.2$ is the path loss exponent, $\beta$ represents the Rician factor, and $\mathbf{H}_t^{\text{LoS}}$ and $\mathbf{H}_t^{\text{NLoS}}$ capture the line-of-sight (LoS) and non-LoS (NLoS) components, respectively. The LoS component $\mathbf{H}_t^{\text{LoS}}$ is given by the product of the steering vector of the RIS and the conjugate transposed steering vector of the BS. As to $\mathbf{H}_t^{\text{NLoS}}$, each entry follows the complex normal distribution with zero mean and unit variance. On the other hand, the channel between the RIS and the $k$th user is subject to path loss and modeled by the semi-correlated NLOS Rayleigh flat fading channel~\cite{Wang2006,Alkhaled2016}, where the fading is correlated at the RIS side but uncorrelated at the users side. Users are uniformly located in a fan-shape area with the radius of 12.5 m and all users are at least 10 m far from the BS, as shown in Fig.~\ref{fig: system-model}. The path loss exponents between the RIS and all the user are equal with $\alpha_{RU} = 2.8$. The rest parameters are listed in Table~\ref{tab: parameters}.

\begin{table}
\centering
\caption{SIMULATION PARAMETERS}
\label{tab: parameters}
\begin{tabular}{ |c|c|c|c| }
\hline
\textbf{Parameter} & \textbf{Assumption} \\
\hline
Bandwidth $B$ & 10 MHz \\
\hline
Noise power $\sigma_{n}$ & -174 dBm/Hz \\
\hline
Number of BS antennas $M$ & 32 \\
\hline
Rician factor $\beta$ & 5 \\
\hline
Total transmit power $P_t$ & 40 dBm \\
\hline
Discrete levels of phase shifts $L$ & 16 \\
\hline
Number of users $K$ & 30 \\
\hline 
RIS configuration overhead $t_p$ & $1\%$ \\
\hline
\end{tabular}
\end{table}

\begin{figure}[!t]
\centering
{
\includegraphics[width=0.87\linewidth]{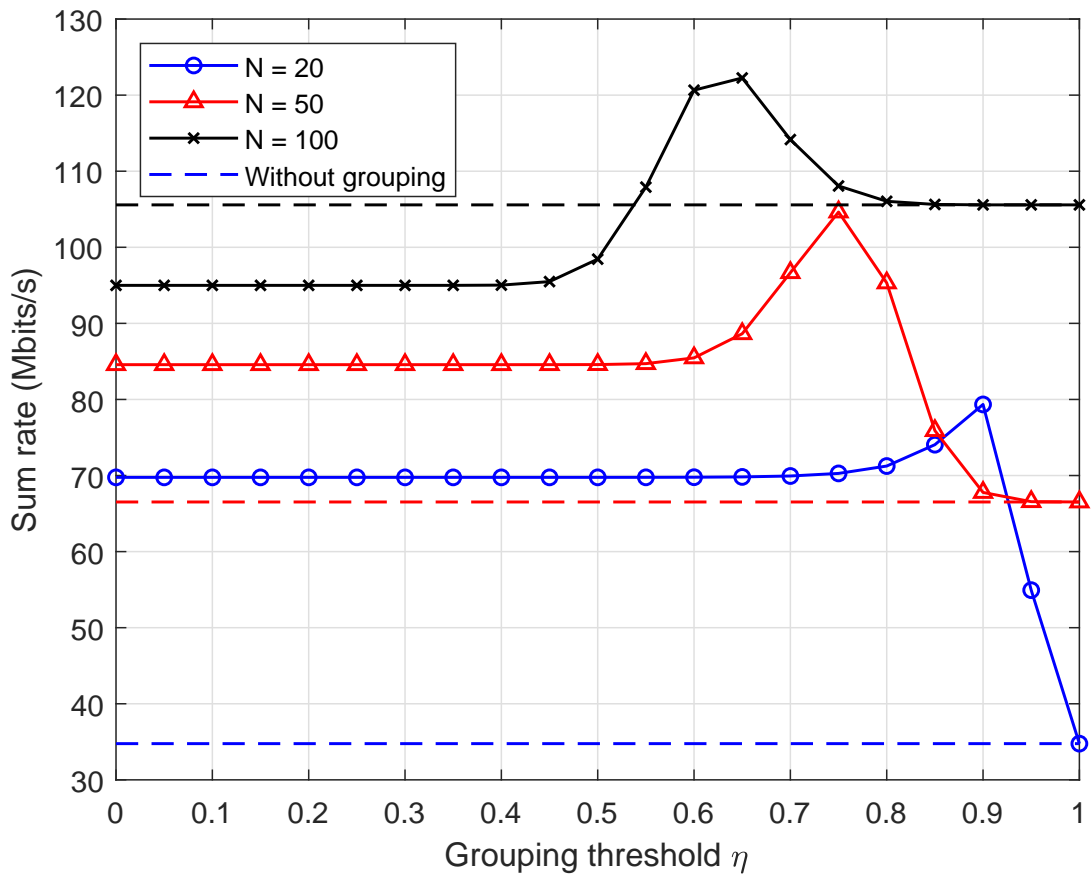}
}
\caption{Sum rate $R_{\text{sum}}$ versus grouping threshold $\eta$ for different number of RIS elements $N$ in the unified design.}
\label{fig: optimal-grouping-threshold}
\end{figure}

Fig.~\ref{fig: optimal-grouping-threshold} plots $R_{\text{sum}}$ versus the grouping threshold $\eta$ for RIS elements $N=20, 50$, and $100$. The dashed lines represent the results when all users are served simultaneously without grouping. A general trend is that $R_{\text{sum}}$ first increases with $\eta$ and then decreases. When $\eta$ is small, the condition $\rho_{i,j} > \eta$ is easier to be met and thus users tend to be separated into different groups, each with a small number of users. 
With more groups created, each group will have less time resources that limits the sum rate. As $\eta$ increases, the trend reverses and there exists a threshold value $\eta$ that maximizes the sum rate. The optimal threshold depends on $N$. From the figure, the optimal threshold value is around 0.9 when $N=20$ and it is about 0.65 when $N=100$. 

\begin{figure}[!t]
\centering
{
\includegraphics[width=0.87\linewidth]{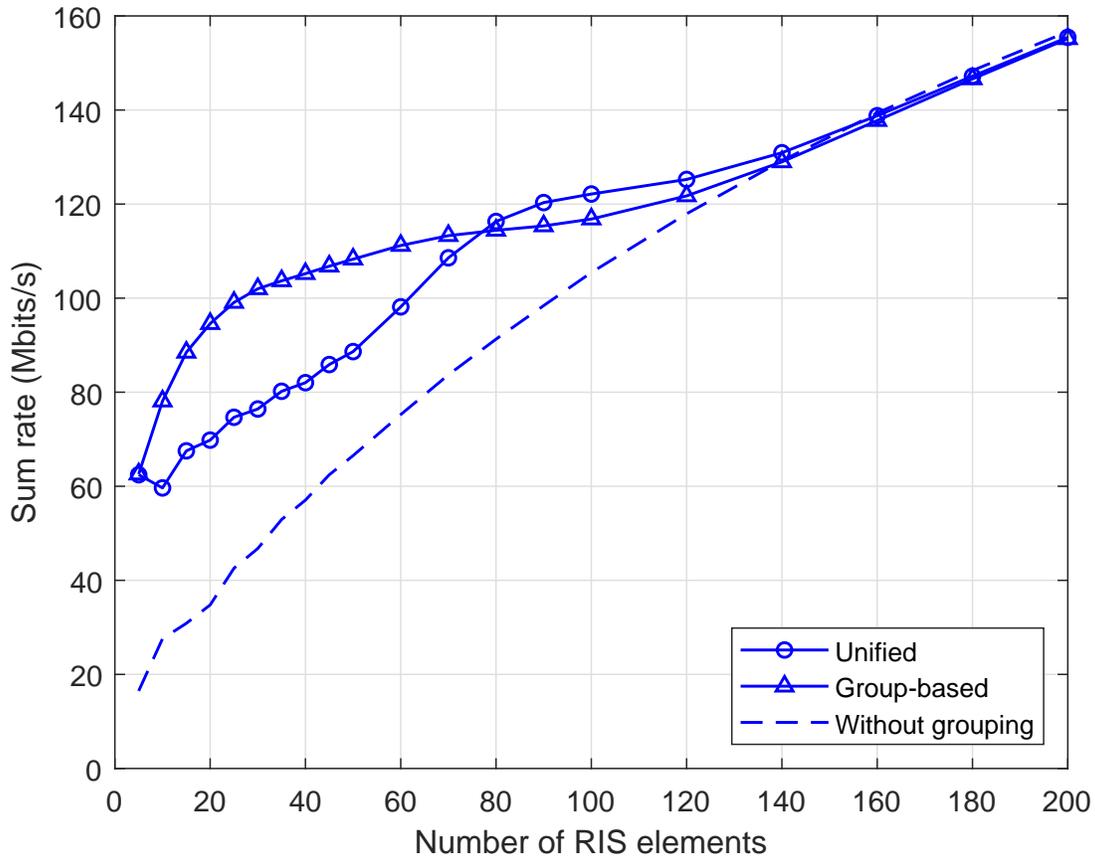}
}
\caption{Sum rate $R_\text{sum}$ versus number of RIS elements $N$ in the group-based design and the unified design for $\eta=0.65$.}
\label{fig: number-of-RIS-elements}
\end{figure}

Next, we show the sum rate of different beamforming strategies, including the group-based design (\ref{eq: beamforming-pair-dependent}) and the unified design (\ref{eq: beamforming-pair-independent}) as a function of the RIS elements $N$ in Fig.~\ref{fig: number-of-RIS-elements}. Here, we fix the threshold $\eta=0.65$ and the result without grouping is also included for comparison. It can be seen that when $N>140$, all the three curves overlap because user grouping is nearly ineffective when $N$ is large as explained in Remark~\ref{rem: correlation}. When $N$ is not so large ($N<140$), both the two proposed beamforming methods significantly outperform the one without grouping, indicating the merit of user grouping. The group-based beamforming design is mostly superior to the unified beamforming design. The latter slightly achieves a higher sum rate than the former because the grouping threshold $\eta$ considered in Fig.~\ref{fig: number-of-RIS-elements} is optimal for the unified beamforming design, as seen from Fig.~\ref{fig: optimal-grouping-threshold}.

\begin{figure}[!t]
\centering
{
\includegraphics[width=0.87\linewidth]{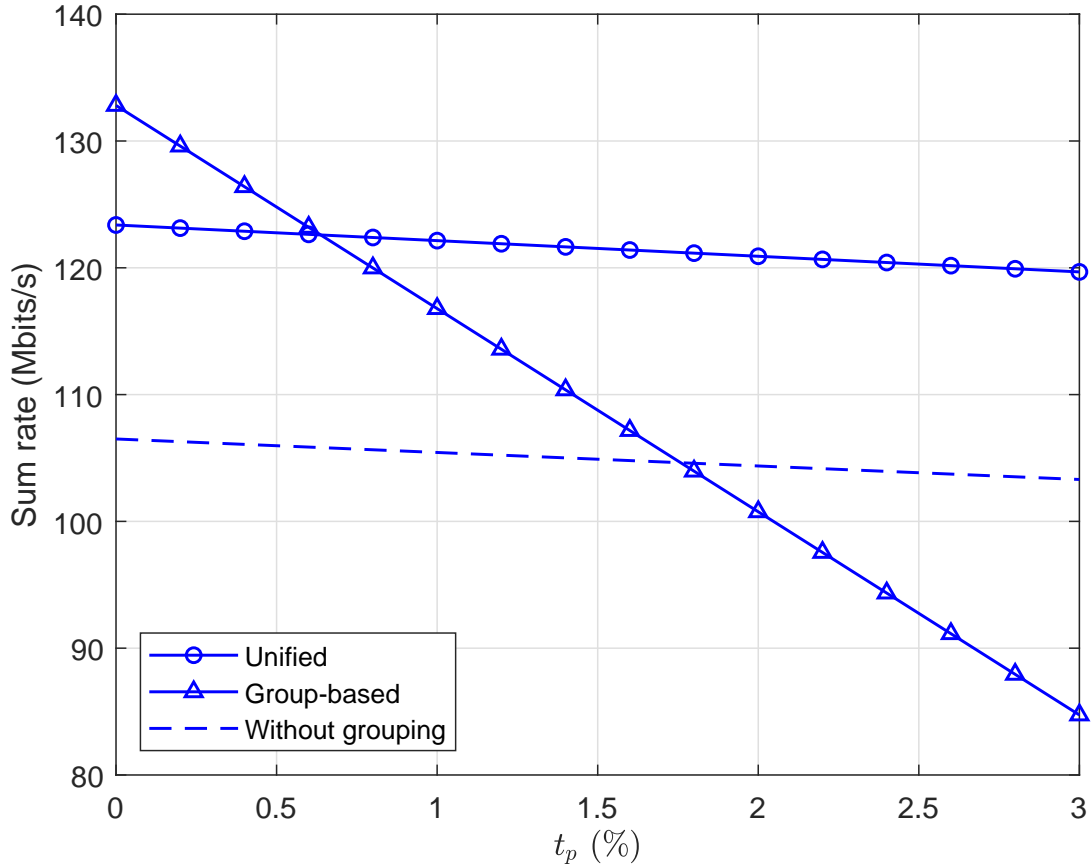}
}
\caption{Sum rate $R_\text{sum}$ versus the RIS configuration overhead $t_p\%$ between the group-based design and the unified design for $\eta=0.65$ and $N=100$.}
\label{fig: adjustment-proportion}
\end{figure}


The impact of RIS configuration time is studied Fig.~\ref{fig: adjustment-proportion} for $N=100$ elements and $\eta=0.65$. Intuitively, when $t_p$ is larger, the sum rates of all the considered schemes degrade. The group-based beamforming is most sensitive to the increase of $t_p$ because it needs more configuration times than other schemes. It is even worse than the case without grouping when $t_p>1.8\%$. The unified beamforming design always outperforms the case without grouping because the former only needs to configure the RIS once per scheduling cycle and it avoids strong correlation by user grouping.

\begin{figure}[!t]
\centering
{
\includegraphics[width=0.87\linewidth]{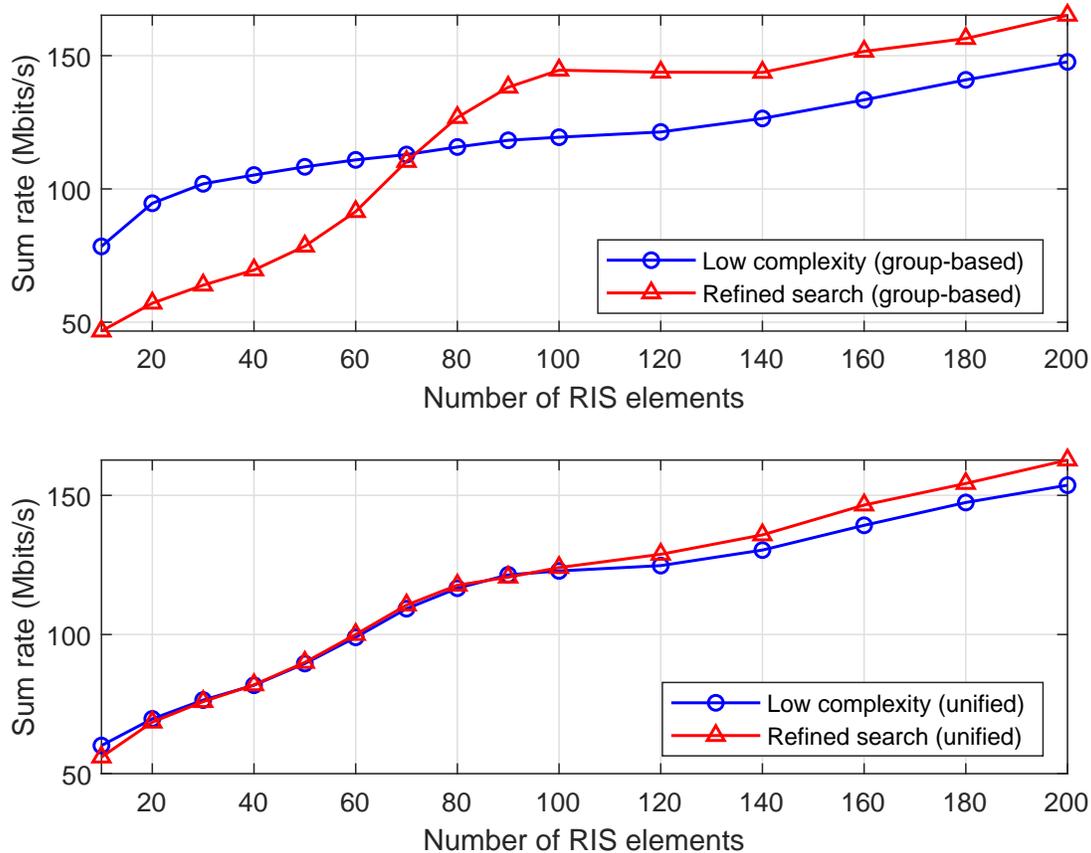}
}
\caption{Comparison of sum rate $R_\text{sum}$ versus number of RIS elements $N$ achieved by the proposed method and AO.}
\label{fig: comparison-beamforming methods}
\end{figure}

In the proposed beamforming design, the beamforming pair is found by a one-dimensional search. In Fig.~\ref{fig: comparison-beamforming methods}, we compare the sum rate achieved by the proposed low-complexity beamforming design with that using multiple searches with refinement as in~\cite{Zheng2020a}, which is indicated as ``Refined search'' (RS) in the figure. Specifically, the RIS is divided into $S$ sub-surfaces and all the reflecting elements in the sub-surface share a common reflection coefficient. The phase shift of one sub-surface is determined by solving (\ref{eq: beamforming-pair-dependent}) or 
(\ref{eq: beamforming-pair-independent}) while fixing the phase shifts of other $S-1$ sub-surfaces. The above procedure iterates until the convergence is reached. The sub-surface structure simplifies the searching complexity for the phase shifts compared to the element-wise optimization. Here we fix $N=100$, $S=5$, and the optimal grouping threshold found through numerical search is applied. First, we focus on the group-based design (top figure of Fig.~\ref{fig: comparison-beamforming methods}). One can see that when $N <70$, the proposed low-complexity scheme achieves a higher sum rate than RS. In this region, the adaptive user grouping algorithm separates correlated users into different groups and each group has only one or two users with low spatial correlations. Thus the sum rate can be better improved by focusing the reflected signal to a specific direction using (\ref{eq: ris-phase-vector}). Different from the proposed method, RS will construct multiple spatial beams for each group and thus it does not perform well when the number of users per group is small. As $N$ increases, the spatial correlation reduces, resulting in more users within the same group as mentioned in Remark~\ref{rem: correlation}. Consequently, having multiple beams optimized by RS is superior than the low-complexity scheme. It is also revealed from the figure that when $N=200$, the proposed method only encounters a minor sum rate loss ($\approx 6\%$) compared to RS yet its complexity is much lower than RS as listed in Table~\ref{tab: complexity}. Next, we observe the unified design (bottom figure of Fig.~\ref{fig: comparison-beamforming methods}). It can be seen that the performance of the unified method is very close to that of RS even when $N$ is large.

\section{Conclusion}\label{sec: conclusion}
This paper aims to simplify the beamforming design for the RIS-assisted multi-user system. 
Simulation results reveal a few key findings as summarized below. i) The proposed group-based beamforming design is advantageous when the number of RIS elements is not very large ($N<70$) and it encounters a small sum rate loss when $N>70$ yet its complexity is much lower than AO-based RIS phase shift refinement. ii) The performance of the group-based and the unified beamforming design can be maximized by optimizing the grouping threshold, which depends on the number of RIS elements and users. iii) The time overhead for configuring RIS phase shifts dramatically affects the sum rate. Even with a small time overhead, say $1\%$ per scheduling cycle, the unified design that configures the RIS once per scheduling cycle performs better than the group-based design. It is found that the optimal grouping threshold plays a vital role in the RIS-assisted multi-user system and thus deserves further study.

\bibliographystyle{IEEEtran}
\bibliography{IEEEabrv,ref_new}

\end{document}